\documentclass[12pt,fleqn]{article}

\usepackage{amsmath, amssymb,graphicx}
\newcommand \be{\begin{equation}}
\newcommand \ee{\end{equation}}
\newcommand \bea{\begin{eqnarray}}
\newcommand \eea{\end{eqnarray}}
\newcommand \bee{\begin{equation}}
\newcommand \eee{\end{equation}}

\newcommand\TL{\hfil$\displaystyle{##}$}
\newcommand\TR{$\displaystyle{{}##}$\hfil}
\newcommand\TC{\hfil$\displaystyle{##}$\hfil}

\def\seqalign#1#2{\vcenter{\openup1\jot
  \halign{\strut #1\cr #2 \cr}}}
\def\lbldef#1#2{\expandafter\gdef\csname #1\endcsname {#2}}
\newcommand{\eqn}[3][]{\lbldef{#2}{(\ref{#2})}%
\def\@eqnstyle{#1}%
\ifx\@eqnstyle\@empty%
\begin{equation} \eqalign{#3} \label{#2} \end{equation}%
\else%
\begin{equation} \seqalign{\span\TC}{#3} \label{#2} \end{equation}%
\fi}
\def\eqalign#1{\vcenter{\openup1\jot
    \halign{\strut\span\TL & \span\TR\cr #1 \cr
   }}}

\def\p2{{p \over 2}}
\def\half{{1\over 2}}
\def\nref#1{(\ref{#1})}
\def\htheta{{\hat \theta}}

\input epsf

\begin{document}


\thispagestyle{empty}
\renewcommand{\thefootnote}{\fnsymbol{footnote}}

{\hfill \parbox{4cm}{ hep-th/0604135 \\
  }}

\bigskip\bigskip

\begin{center} \noindent \Large \bf
Giant Magnons
\end{center}

\bigskip\bigskip\bigskip

\centerline{ \normalsize \bf Diego  M. Hofman$^a$ and Juan
Maldacena$^{b}$\footnote[1]{\noindent \tt dhofman@Princeton.edu ,
malda@ias.edu} }

\bigskip
\bigskip\bigskip

\centerline{$^a$ \it Joseph Henry Laboratories, Princeton
University, Princeton, NJ 08544, USA}
\bigskip
\centerline{$^d$ \it Institute for Advanced Study, Princeton NJ
08540, USA.}
\bigskip\bigskip

\bigskip\bigskip

\renewcommand{\thefootnote}{\arabic{footnote}}

\centerline{\bf \small Abstract}
\medskip

{\small Studies of ${\cal N}=4$ super Yang Mills operators with
large R-charge have shown that, in the planar limit,  the problem
of computing their dimensions can be viewed as a certain spin
chain. These spin chains have fundamental ``magnon'' excitations
which obey a  dispersion relation that is periodic in the momentum
of the magnons. This result for the dispersion relation was also
shown to hold at arbitrary 't Hooft coupling. Here we identify
these magnons  on the string theory side and we show how to
reconcile a periodic dispersion relation with the  continuum
worldsheet description. The crucial idea is that the momentum is
interpreted in the string theory side as a certain geometrical
angle. We use these results to compute the energy of a spinning
string. We also show that the symmetries that determine the
dispersion relation and that constrain the S-matrix are the same
in the gauge theory and the string theory. We compute the overall
S-matrix at large 't Hooft coupling using the string description
and we find that it agrees with an earlier conjecture. We also
find an infinite number of two magnon bound states at strong
coupling, while at weak coupling this number is finite.
 }

\newpage

\section{Introduction}
\label{INTRODUCTION}

String theory in $AdS_5 \times S^5$ should be dual to ${\cal N}=4$
Yang Mills \cite{jm,gkp,wittenhol}. The spectrum of string states
should be the same as the spectrum of operators in the Yang Mills
theory. One interesting class of operators are those that have
very large charges \cite{bmn}. In particular, we consider
operators where one of the SO(6) charges, $J$, is taken to
infinity. We study states which have finite $E-J$. The state with
$E-J=0$ corresponds to a long chain (or string) of $Z$s, namely to
the operator $Tr[Z^J]$. We can also consider a finite number of
other fields $W$ that propagate along this chain of $Z$s. In other
words we consider operators of the form
  \be \label{operdef}
  O_p \sim  \sum_l e^{i
 {l p }} ( \cdots ZZZWZZZ \cdots)
 \ee
 where the field $W$ is inserted at position $l$ along the chain.
  On the gauge theory side the problem of diagonalizing the planar Hamiltonian
   reduces to a type of
spin chain \cite{minzar,beisertoneloop}, see \cite{otherreview}
for reviews and further references. In this context the
impurities, $W$, are ``magnons'' that move along the chain.

Using supersymmetry, it was shown that these excitations have a
dispersion relation of the form \cite{beiserts}
 \be \label{disprel}
 E-J = \sqrt{ 1 + { \lambda \over \pi^2} \sin^2{p\over 2} }
 \ee
Note that the periodicity in $p$ comes from the discreteness of
the spin chain.
 The large 't Hooft coupling limit of this result is
  \be \label{larget}
  E- J = { \sqrt{\lambda} \over \pi} \left| \sin{p\over 2} \right|
  \ee
  Since this is a strong coupling result, it should be possible to
  reproduce it on the string theory side.
  At first sight it would seem that such a dispersion relation
  would require the string worldsheet to be discrete. In fact,
  this is not the case.
 We will show how to recover \nref{larget} on the
 string theory side with the usual strings moving in $AdS_5 \times S^5$.
The key element is that $p$ becomes a geometrical angle which will
explain the periodic  result. Thus we are able to identify the
elementary excitations of the spin chain on the string theory side
in an explicit fashion. The identification of these magnons allows
us to explain, from the gauge theory side, the energy spectrum of
the string spinning on $S^5$ which was considered in
\cite{gkpspin}.

We will discuss the presence of extra central charges in the
supersymmetry algebra which match the gauge theory analysis in
\cite{beiserts}. Having shown that the two algebras match, then
the full result \nref{disprel} follows. Moreover, as shown in
\cite{beiserts} the symmetry algebra constrains the $2 \to 2 $ $
S$ matrix for these excitations up to an overall phase. This $S$
matrix is the asymptotic S-matrix discussed in \cite{staudachers}.
 It should be emphasized that these magnons
are the fundamental degrees of freedom
 in terms of which we can construct all other states of the system.
  Integrability     \cite{Mandal,bpr,frolovint} implies
 that the scattering of these excitations is dispersionless.
 We check that this is indeed the case classically  and we
  compute the classical
 time delay for the scattering process. This determines the
 large 't Hooft coupling limit of the scattering phase. The final
 result agrees with the large $\lambda$ limit of
 \cite{stringbethe}. This is done by exploiting a connection with
 the sine Gordon model \cite{pohlmeyer,mikhailov}. We also find
 that at strong coupling there is an infinite number of bound
 states of two magnons. These bound states have more energy than
 the energies of the individual magnons.

 This article is organized as follows. In section 2 we explain
 the string theory picture for the magnons.
  We start by
 defining a particular limit that lets us isolate the states we
 are interested in. We continue with a review of gauge theory
 results in this limit. Then we find solutions of the classical
 sigma model action which describe magnons. We present these
 solutions in various coordinate systems. We also explain how the
 symmetry algebra is enhanced by the appearance of extra central
 charges, as in the gauge theory side \cite{beiserts}. Finally, we
 end by applying these results to the computation of the energy of
 a spinning string configuration in $S^5$ considered in \cite{gkp}.
In section 3 we compute the S-matrix at strong coupling and we
analyze the bound state spectrum.

In appendix A we give some details on the supersymmetry algebra.
In appendix B give some more details on the spectrum of bound
states.

 \section{ Elementary excitations on an infinite string}

\subsection{ A large $J$ limit}

 Let us start by specifying the limit that we
 are going to consider. We will first take the ordinary 't Hooft
 limit. Thus we will consider free strings in $AdS_5 \times S^5$ and planar diagrams
 in the gauge theory.
 We then pick a generator $J = J_{56} \subset so(6)$ and consider
 the limit when $J$ is very large. We will consider states with
 energies $E $ (or operators with conformal dimension $\Delta =E$)
 which are such that $E-J$ stays finite in the limit.
   We keep the 't Hooft coupling
 $\lambda \equiv g^2 N $
 fixed. This limit can be considered both on the gauge theory and
 the string theory sides and we can interpolate between them by varying the
 't Hooft coupling  after having
 taken the large $J$  limit. In addition, when we consider an
 excitation we will keep its momentum $p$ fixed.
 In summary, the limit that we are considering is
 \bea
  J & \to & \infty ~,~~~~~~~~ \lambda = g^2 N = {\rm fixed}
  \\
  p &=& {\rm fixed } ~,~~~~~E-J = {\rm fixed }
  \eea
 This differs from the plane wave   limit \cite{bmn} in two ways. First,
 here we are keeping $\lambda$ fixed, while in \cite{bmn} it was
 taken to infinity. Secondly, here we are keeping
 $p$ fixed, while in \cite{bmn} $n = p J$ was kept fixed.

 One nice feature of this limit is that it decouples quantum
 effects, which are characterized by $\lambda$, from finite $J$ effects, or
 finite volume effects on the string worldsheet\footnote{ The importance of
 decoupling these two effects was emphasized in
 \cite{mannpolch}.}.

 Also, in this limit, we can forget about the momentum constraint
 and think about single particle excitations with non-zero
 momentum. Of course, when we take $J$ large but finite, we will
 need to reintroduce the momentum constraint.

\subsection{Review of gauge theory results}

In this subsection we will review the derivation of the formula
\nref{disprel}.  This formula could probably have been obtained in
\cite{sz} had they not made a small momentum approximation at the
end. This formula also emerged via perturbative computations
\cite{otherreview}.  A heuristic explanation was given in
\cite{berenstein}, which is very close to the string picture that
we will find below.
  Here we will follow
the treatment in \cite{beiserts} which exploits some interesting
features of the symmetries of the problem.

The ground state of the system, the state with $E-J=0$, preserves
16 supersymmetries\footnote{More precisely it is annihilated by 16
+ 8, but the last 8 act non-linearly on the excitations, they
correspond to fermionic impurity annihilation operators with zero
momentum.}. These supercharges, which have $E-J=0$,  act linearly
on the impurities or magnons.   They transform as $ (2,1,2,1) +
(1,2,1,2) $ under $SU(2)_{S^5 , L} \times SU(2)_{S^5, R} \times
SU(2)_{AdS_5 , L} \times SU(2)_{AdS_5 , R} $ where the various
$SU(2)$ groups corresponds to the rotations in $AdS_5$ and $S^5$
which leave $Z$ invariant. These supercharges are the odd
generators of two  $SU(2|2)$ groups\footnote{Note that they are
not $PSU(2|2)$ groups.}. The energy $\epsilon \equiv E-J$ is the
(non-compact) $U(1)$ generator in each of the two $SU(2|2)$
supergroups. In other words, the two $U(1)$s of the two $SU(2|2)$
groups are identified. A single impurity with $p=0$ transforms in
the smallest BPS representation of these two supergroups. In
total, the representation has $8  $ bosons plus $8$ fermions. This
representation is BPS because its energy is $\epsilon = E-J =1$
which follows from the BPS bound that links the energy to the
$SU(2)^4$ charges of the excitations. Let us now consider
excitations with small momentum $p$. At small $p$ we can view the
dispersion relation as that of a relativistic theory. Note that as
we add a small  momentum, the energy becomes higher but we still
expect to have 8 bosons plus 8 fermions and not  more, as it would
be the case for representations of $SU(2|2)^2$ with $\epsilon >1$.
What happens is that the momentum appears in the right hand side
of the supersymmetry algebra. This ensures that the representation
is still BPS.  In fact, for finite $p$ there are two central
charges \cite{beiserts}. These extra generators add or remove $Z$s
to the left or right of the excitation and they originate from the
commutator terms in the supersymmetry transformation laws, namely
terms like $\delta W \sim  \psi + [Z, \chi] $, see
\cite{beiserts}.
 These extra central charges are zero for
physical states with finite $J$ since we will impose the momentum
constraint.

The full final algebra has thus three ``central'' generators in
the right hand side, they are the energy $\epsilon$ and two extra
charges which we call $k^1, k^2$. Together with the energy these
charges can be viewed as the three momenta $k^\mu$ of a 2+1
dimensional Poincare superalgebra. This is the same as the 2+1
dimensional Poincare superalgebra recently studied in
\cite{lm,seiberg}, we will see below that this is not a
coincidence. Notice that the Lorentz generators are an outer
automorphism of this algebra but they are not  a symmetry of the
problem we are considering. See appendix A for a more detailed
discussion of the algebra.

As explained in \cite{beiserts} the expression for the ``momenta''
is $k^1 + i k^2 = h(\lambda) ( e^{ip}-1)$ and similarly for the
complex conjugate. This then implies that we have the formula
 \be
 \label{enerrelat}
 E - J = k^0 = \sqrt{ 1 + |k_1 + i k_2|^2 } =
 \sqrt{ 1 + f(\lambda) \sin^2{p \over 2} }
 \ee
 The function $f(\lambda)$ is not determined by this symmetry
 argument.
We know that  $f(\lambda) = { \lambda \over \pi }$ up to three
loops in the gauge theory \cite{minzar,3loops}  and that it is
also the same at strong coupling (where it was checked at small
momenta in \cite{bmn}). \cite{sz} claims to show it is exactly $f
=  {\lambda \over \pi } $ for all values of the coupling, but we
do not fully understand the argument\footnote{ It is not clear to
us why in equation (10) in \cite{sz} we could not have a function
of $\lambda$ in the right hand side.}.

In the plane wave matrix model \cite{bmn,plefkamm} one can also
use the symmetry algebra to determine a dispersion relation as in
\nref{enerrelat} and the function $f(\lambda)$ is nontrivial. More
precisely, large $J$ states in the plane wave matrix model have an
 $SU(2|2)$ group (extended by the central charges
 to a 2+1 Poincare superalgebra) that acts
 on the impurities.

 The conclusion is that
elementary excitations moving along the string are BPS under the
16 supersymmetries that are linearly realized. Supersymmetry then
ensures that we can compute the precise mass formula once we know
the expression for the central charges.

\subsection{String theory description at large $\lambda$}

 We will now give the description of the elementary impurities or
 elementary magnons at large $\lambda$ from the string theory side.
 In this regime  we can trust the classical approximation to the string
 sigma model in $AdS_5 \times S^5$.

\begin{figure}[htb]
\begin{center}
\epsfxsize=4in\leavevmode\epsfbox{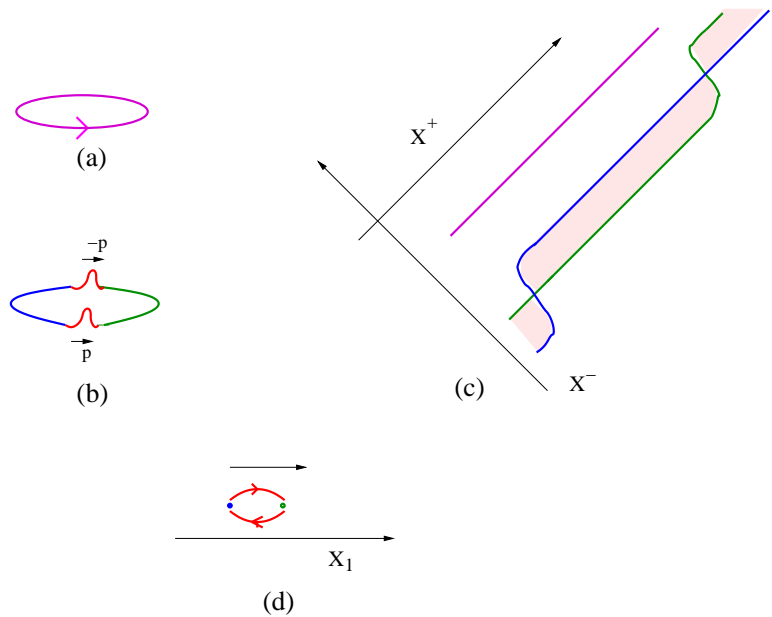}
\end{center}
\caption{ Localized excitations propagating along the flat space
string worldsheet in light cone gauge.  (a) Worldsheet picture of
the light cone ground state, with $P_+=0$. (b) Worldsheet picture
of two localized excitations with opposite momenta propagating
along the string.  (c) Spacetime description of the configurations
in (a) and (b). The configuration in (a) gives a straight line at
a constant $X^-$. The configuration in (b) gives two straight
lines at constant $X^-$ when the localized excitations are
separated on the worldsheet. When the two excitations in (b) cross
each other the lines move in $X^-$.
  (d) Snapshot of the spacetime configuration in (b), (c) at a given time $t$.
   } \label{lightlike}
\end{figure}

 In order to understand the solutions that we are going to study,
 it is convenient  to consider first a string in flat space. We
 choose light cone gauge, with $X^+=\tau$,
  and consider a string with large $P_-$.
 The solution with $P_+=0$ corresponds to a lightlike trajectory
 with $X^- =$constant, see figure \ref{lightlike}(a,c).
  Now suppose that we put two localized
 excitations carrying worldsheet momentum $p$ and $-p$ respectively. Let us
 suppose that at some instant of time these are on opposite points
 of the worldsheet spatial circle, see figure \ref{lightlike}(b).
 We want to understand the
 spacetime description of such states.
   It is clear that the
 region to the left of the excitations and the region to the right
 will be mapped to the same lightlike trajectories with $X^-
 =$constant that we considered before. The important point is that
 these two trajectories sit at different values of $X^-$.  This can be seen by writing
 the Virasoro constraint as
 \be
  \partial_\sigma X^- = 2 \pi \alpha' T_{\tau \sigma } ~,~~~~~~~~\Delta X^- =
   2 \pi \alpha' \int
  d \sigma T_{\tau \sigma} =  2 \pi \alpha' p
  \ee
 where $T_{\tau \sigma}$ is the worldsheet
 stress tensor of the transverse
 excitations and we have integrated across the region where the
 excitation with momentum $p$ is localized.
 Thus the final spacetime picture is that we have two particles
 that move along lightlike trajectories that are joined by a
 string. At a given time $t$ the two
 particles move at the speed of light
  separated by $\Delta X^1|_t = \Delta X^-|_{X^+} = 2\pi \alpha' p
  $ and joined by a string, see figure \ref{lightlike}(d).
  Of course, the string takes momentum
 from the leading particle and transfers it to the trailing
 one. On the worldsheet this corresponds to the two localized
 excitations moving toward each other. As the worldsheet
 excitations pass through each other the trailing particle becomes
 the leading one, see figure \ref{lightlike}(c). For a closed string
 $X^-$ should be periodic, which leads to the momentum constraint
 $p_{total} =0$.

\begin{figure}[htb]
\begin{center}
\epsfxsize=4in\leavevmode\epsfbox{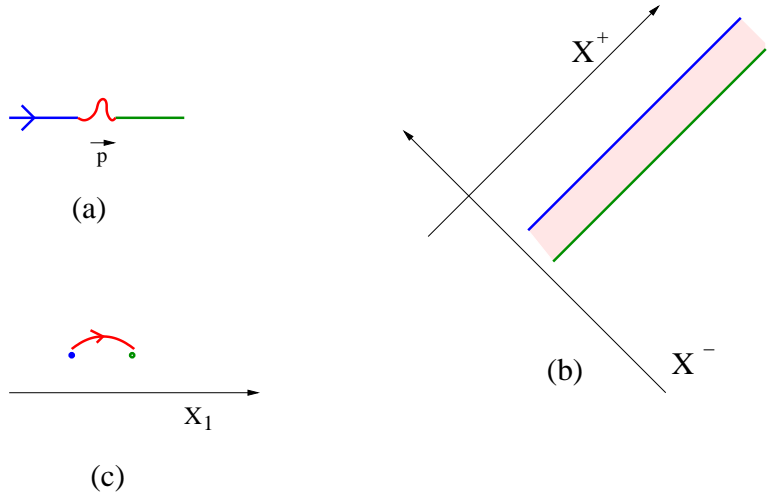}
\end{center}
\caption{ Localized excitations propagating on an infinite string.
(a) Worldsheet picture of a localized excitation propagating along
the string. (b) Spacetime behavior of the state in  lightcone
coordinates. We have two lightlike lines with a string stretching
between them.  (c) Snapshot of the state at a given time. The
configuration moves to the right at the speed of light. }
\label{lightlike2}
\end{figure}

 In the limit of an infinite string, or infinite $P_-$, we can
 consider a single excitation with momentum $p$ along an infinite
 string. Then the spacetime picture will be that of figure
 \ref{lightlike2}
 where we have two lightlike trajectories, each carrying infinite
 $P_-$, separated by $\Delta X^- \sim p$ which are joined by a
 string. There is some $P_-$ being transferred from the
 first to the second. But since $P_-$ was infinite  this can
 continue happening for ever\footnote{ As a side remark, notice that these
 lightlike trajectories look a bit like light-like D-branes, which could be viewed
 as small giant gravitons in the $AdS_5 \times S^5$ case.
  In this paper we take the 't Hooft limit before the large
 $J$ limit so we can ignore giant gravitons. But it might be worth
 exploring this
 further. Strings ending in giant gravitons were recently studied in \cite{BCV}}.
 The precise shape of the string that joins the two points depends
 on the precise set of transverse excitations that are carrying
 momentum $p$.

 Armed with the intuition from the flat space case, we can now go
 back to the $AdS_5 \times S^5$ case.
 We write the metric of $S^5$ as
 \be
 ds^2 = \sin^2 \theta d\varphi^2 + d\theta^2 + \cos^2 \theta d\Omega_3^2
 \ee
 where $\varphi$ is the coordinate that is shifted by $J$.
The string ground state, with $E-J=0$, corresponds to a lightlike
trajectory   that moves along $\varphi$, with $\varphi - t
$=constant, that sits at $\theta=\pi/2$ and
 at the origin of the spatial directions of $AdS_5$.

\begin{figure}[htb]
\begin{center}
\epsfxsize=1in\leavevmode\epsfbox{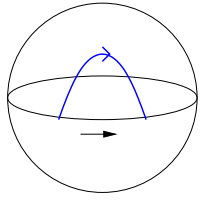}
\end{center}
\caption{Giant magnon solution to the classical equations. The
momentum of the state is given by the angular distance between the
endpoints of the string. We depicted a configuration with $ 0 < p
< \pi$. A configuration with negative momentum would look the same
except that the orientation of the string would be reversed.
  The string endpoints are on the equator and the move at the speed of
  light.
 } \label{spheresol}
\end{figure}

 We can find the solution we are looking for in various ways.
 We are interested in finding the configuration
 which carries momentum $p$ with least amount of energy $\epsilon
 = E-J$. For the moment let us find a solution with the expected
 properties and we will later show that it has the minimum amount
 of energy for fixed $p$.
 We first pick a pair of antipodal points on $S^3$ so that, together
 with the coordinate $\theta$ and $\varphi$ they form an $S^2$.
 After we include time, the motion takes place in $R\times
 S^2$. We can now write the Nambu action choosing the
 parametrization
 \be
 t= \tau ~,~~~~~~~~~\varphi - t = \varphi'
 \ee
 and we consider a configuration where $\theta$ is independent of
 $\tau$. We then find that the action reduces to
 \be
 S = { \sqrt{\lambda} \over 2 \pi }
 \int dt d\varphi' \sqrt{ \cos^2 \theta { \theta '}^2 + \sin^2 \theta }
 \ee
 It is easy to integrate the equations of motion and we get
 \be \label{nambusol}
 \sin \theta = { \sin \theta_0 \over \cos \varphi' } ~,~~~~~~
 - ( { \pi \over 2} - \theta_0)  \leq \varphi' \leq  { \pi \over 2 } - \theta_0
 \ee
 where $0 \leq  \theta_0 \leq \pi/2$ is an integration constant. See
 figure \ref{spheresol}.
 In these variables the string has finite worldsheet extent, but
 the regions
 near the end points are carrying an infinite amount of
 $J$.
 We see that for this solution the difference in angle between the
 two endpoints of the string at a given time $t$ is
 \be \label{delvar}
 \Delta \varphi' = \Delta \varphi  = 2 ( { \pi \over 2} -
 \theta_0 )
 \ee
 It is also easy to compute the energy
 \be \label{energy}
 E- J = { \sqrt{\lambda} \over \pi} \cos \theta_0 =  { \sqrt{\lambda} \over \pi}
 \sin { \Delta \varphi  \over 2}
 \ee
 We now propose the following identification for the momentum $p$
 \be
  \Delta \label{momid} \varphi  = p
 \ee
 We will later see more evidence for this relation.
Once we use this relation \nref{energy} becomes
 \be \label{finenergy}
  E-J = { \sqrt{\lambda} \over \pi}  \left| \sin {p \over 2} \right|
 \ee
 in perfect
agreement with the large $\lambda$ limit \nref{larget} of the
gauge theory result \nref{disprel}. The sign of $p$ is related to
the orientation of the string. In other words, $\Delta \varphi$ is
  the angular position of the endpoint of the string minus that
of the starting point and it can be negative.

In order to make a more  direct comparison with the gauge
 theory it is useful to pick a gauge on the worldsheet
 in such a way that, for the string ground state (with $E-J=0$),
 the density of $J$ is constant\footnote{Note that we only require $J$ to be
 constant away from the excitations, it could or could not be constant in the
 regions where $E-J >0$.}. There are various ways of doing
 this. One specific choice would be the light cone gauge
 introduced in \cite{frolovetal}.
 We will now do something a bit different which can be done easily
 for strings on $R \times S^2$ and which will turn out useful for our later
 purposes.  This consists in choosing
 conformal gauge and setting $t=\tau$. $x$ labels the   worldsheet
 spatial coordinate.
 In this gauge the previous
 solution takes the form
 \bea \label{confsol}
 \cos \theta &=& { \cos \theta_0 \over \cosh \left[ { x - \sin \theta_0 t
 \over \cos \theta_0 }  \right]} ={ \sin {p\over 2}  \over \cosh \left[ { x - \cos {p \over 2} t
 \over \sin { p \over 2 }  } \right] }
 \\
  \tan (\varphi - t) &=& \cot \theta_0  \,  \tanh \left[  { x - \sin \theta_0 t
 \over \cos \theta_0  } \right] = \tan \p2  \,  \tanh \left[  { x - \cos {p \over 2} t
 \over \sin { p \over 2 }  } \right]    \notag
 \eea
 where we used \nref{momid}.
 In this case we see that the range of $x$ is infinite. These
 coordinates have the property that the density of $J$ away from
 the excitation is constant. This property allows us a to make an identification
 between the coordinate $x$ and the position $l$ (see \nref{operdef})
 along the chain in the gauge theory.
 More precisely, we compute  the density of $J$ per unit $x$ in
 order to relate $l$ and $x$
 \be \label{xandl}
  { d J \over d x} = { \sqrt{\lambda } \over 2 \pi} ~~~~~~~{\rm or
  }~~~~~~~
  d l = { \sqrt{\lambda } \over 2 \pi} dx
  \ee
 This relation allows us  to check the identification of the
 momentum \nref{momid} since the relation between energy and momentum
 \nref{larget} determines the velocity in the gauge theory
 through the usual formula
 \be \label{velgauge}
 v_{gauge} = { dl \over dt} = { d \epsilon(p) \over d p }  = {\sqrt{\lambda} \over 2 \pi}
 \cos \p2  ~,~~~~~~~~~{\rm for}~~~~p>0
 \ee
 On the other hand we see from \nref{confsol} that the velocity is
 \be \label{velstring}
 v_{string} = { d x \over d t} = \sin \theta_0 = \cos { \Delta
 \varphi \over 2}
 \ee
 We see that after taking into account \nref{xandl} the two
 velocities become identical if we make the identification
 \nref{momid}.

\begin{figure}[htb]
\begin{center}
 \epsfxsize=4in\leavevmode\epsfbox{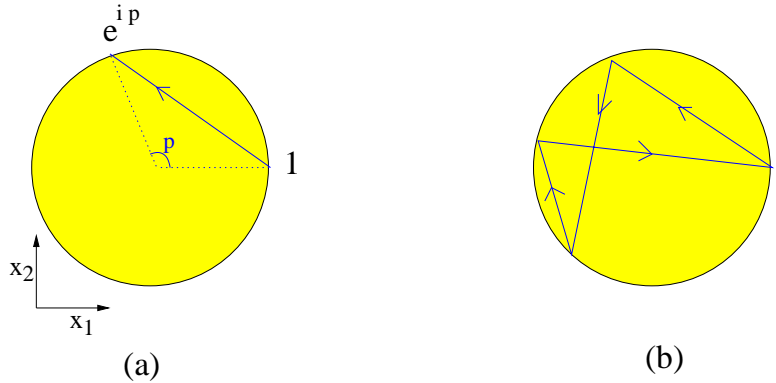}
\end{center}
\caption{Giant magnons in LLM  coordinates \nref{llmcoord}. (a) A
giant magnon solution looks like a straight stretched string. It's
momentum $p$ is  the angle subtended on the circle. $k_1$ and
$k_2$ are the projections of the string along the directions $\hat
1$ and $\hat 2$. The direction of the string gives the phase of
$k_1 + i k_2$, while its length gives the absolute value of the
same quantity. (b) A closed string state built of magnons that are
well separated on the worldsheet. Notice that the total central
charges $k_1,~ k_2$ vanish. Similarly the total angle subtended by
the string, which is the total momentum $p_{total}$ also vanishes
modulo $2 \pi$.
 } \label{circle}
\end{figure}

 The solution becomes simpler if expressed in terms of the
 coordinates introduced in \cite{llm}. Those coordinates were
 specially adapted to describe 1/2 BPS states which carry charge
 $J$. So it is not surprising that they are also useful for
 describing   small excitations around such states.
 The $AdS_5 \times S^5$  metric in those coordinates is a fibration of $R$, characterizing
 the time direction, and two $S^3$s (coming form $AdS_5$ and $S^5$) over a three dimensional
 space characterized by coordinates $x_1,x_2, y$. The plane $y=0$ is special because one of
 the two 3-spheres shrinks to zero size in a smooth way. Thus the plane $y=0$ is divided into
 regions (or ``droplets'') where one or the other $S^3$ shrinks to zero size. The $AdS_5 \times S^5$ solution
 contains a single circular droplet of radius $R$ where the $S^3$ coming from $AdS_5$ shrinks,
 see figure \ref{circle}. Particles carrying $E-J=0$ live on the boundary of
 the two regions. In fact the circle constituting the boundary of
 the two regions sits at $\theta = \pi/2$ and it is parameterized by
  $\varphi' = \varphi - t$ in previous
 coordinates.
 We will be only interested in the metric on this special plane at
 $y=0$
   which takes the form, for $r<1$,
  \be \label{llmcoord}
  ds^2 =R^2 \left\{ -  (1 -r^2) \left[ dt - { r^2 \over (1-r^2)}
  d\varphi' \right]^2 + { dr^2 + r^2 d{\varphi'}^2 \over (1-r^2 )}
  + (1 - r^2) d\Omega_3^2 + \cdots \right\}
  \ee
 where $r^2 = \sin^2 \theta = x_1^2 + x_2^2$ and the dots remind us that we
 are ignoring the $y$ coordinate and the second sphere,
 which has zero size at $y=0$ for
 $r\leq 1$.

 In these coordinates the
  solution is simply a straight line that joins two points
  of the circle as shown in figure \ref{circle}(a). This can be
  seen from \nref{nambusol}, which can be rewritten as
   \be
  r \cos \varphi' = x_1 = {\rm const}
  \ee
  The energy is
  simply the length of the string measured with the flat metric on
  the plane parameterized by $x_1,x_2$; $ds^2_{flat} = R^2 (dx_1^2 + dx_2^2)$.
 In fact, the picture we are finding here is almost identical to
  the one discussed from the gauge theory point of view in
  \cite{berenstein}\footnote{The difference is that
  \cite{berenstein} considered an $S^5$ in $R^6$ and a string
  stretching between two points on $S^5$ through $R^6$.}.
  If we restrict the arguments in \cite{berenstein} to 1/2 BPS
  states and their excitations we can see how this picture emerges
  from the gauge theory point of view. Namely,
   we first diagonalize the matrix $Z$ in
  terms of eigenvalues. Then the impurity is  an off
  diagonal element of a second matrix $W$ which joins two
  eigenvalues that are at different points along the circle.
  The energy formula follows from the commutator term $tr|[Z,W]|^2$ in the
  gauge theory, see \cite{berenstein} for
  more details.

  These coordinates are
  very useful for analyzing the symmetries. In particular, we will
  now
  explain the appearance of extra central charges and we will
  match the superalgebra to the one found on the gauge theory side in
 \cite{beiserts}.
  Under general
  considerations we know that the anticommutator of two
  supersymmetries in 10 dimensional supergravity contains
   gauge transformations for the NS-$B_2$ field \cite{schwarz}.
   These act non-trivially on stretched strings.
     In flat space this leads
  to the fact that straight strings are BPS.
   In fact, inserting the  explicit expression of the Killing
  spinors in \cite{llm} into the general formula for the anticommutator
  of two supercharges \cite{schwarz} it is possible to see that
   the
 relevant NS gauge transformations are those with a constant gauge
 parameter $\Lambda_1 , ~\Lambda_2$, $\delta B = d\Lambda $ \footnote{
 The requisite spinor bilinear is closely related to the one in eqn. (A.45) of
 \cite{llm}. Namely, the expression in \cite{schwarz} involves
 terms of the form
 ${\bar \epsilon}_1^* \gamma^\mu \epsilon_2$, which becomes
  $\epsilon^{t} \Gamma^2 \Gamma^\mu
 \epsilon $ in the notation of \cite{llm}.  }.
 Strings that are stretched along the     $\hat 1$ or $\hat 2$
  directions acquire a phase under such gauge transformations.
  Thus these are the central charges that we are after.
    Note that in order to activate these central charges it
  is not necessary to have a compact circle in the geometry.
   In fact,  the string stretched between
  two separated D-branes in flat space is
   BPS for the same reason\footnote{ If we
  think of the string with $J=\infty$ as a lightlike D-brane,
  the analogy becomes
  closer.}.

  Actually,  the  supersymmetry
  algebra is identical to  a supersymmetry algebra in 2+1 dimensions,
  where the string winding charges, $k^1, k^2$, play the role of the spatial
  momenta\footnote{All these statements are independent
  of the shape
   of the droplets in \cite{llm}. This particular statement is easiest
  to see if we consider droplets on a torus  and we perform a T-duality which takes us
  to a 2+1 dimensional Poincare invariant theory (in the limit the original torus is very small).
  This is a theory studied in \cite{llm,seiberg}. This also explains why the supersymmetry
  algebra is the same in the two problems.}.
   See appendix A for more details
  on the algebra.  From the 2+1 dimensional point of
  view  it is a peculiar
  Poincare   super algebra since it has $SO(4)^2$ charges in the
  right hand side of supersymmetry anti-commutators.
   Of course, this is the
  same supersymmetry algebra that appeared in the gauge theory
  discussion \cite{beiserts}. In conclusion, the symmetry algebra
  is exactly the same on both sides. The extra central charges
  are related to string winding charges. We can think of the
  vector given by the stretched string as the two spatial momenta
  $k^1,~k^2$ appearing in the Poincare superalgebra.
  In other words, we can literally think of the stretched string
  in figure \ref{circle} as specifying a vector $k^1,k^2$ of size
  \be \label{momcen}
  k^1 + i k^2 = { R^2 \over 2 \pi \alpha'} ( e^{ip} -1 ) =
  i { \sqrt \lambda \over \pi } e^{i \p2 } \sin \p2
  \ee
 Then the usual relativistic formula for
 the energy implies \nref{enerrelat}, as in the gauge theory.
 Note that the problem we are considering
  does {\it not} have lorentz invariance in 2+1 dimensions. Lorentz invariance is an
  outer automorphism of the algebra, that is useful for analyzing representations of the
  algebra, but it is not an actual symmetry of the theory.
 In particular, in our problem the formula \nref{enerrelat} is not a
 consequence of boost invariance, since boosts are not
 a symmetry\footnote{One might wonder whether  boosts are a hidden symmetry of the string
 sigma model. This is not the case because we can increase $|k|$ without bound by performing
 a boost, while physically we know that $|k|$ is bounded as in \nref{momcen}.}.
  It is a consequence of supersymmetry, it is a BPS
 formula. Note that rotations of $k^1, ~k^2$ are indeed a symmetry
 of the problem and they correspond to rotations of the circle in
 figure \ref{circle}. This is the symmetry generated by $J$.
 Note also that a physical state with large
 but finite $J$ will consist of several magnons but the
 configurations should be such that we end up with a closed
 string, see figure \ref{circle}(b) .
 Thus, for ordinary closed strings   the total value of the central
 charges is zero, since there is no net string winding. This
 implies, in particular, that for a closed finite $J$ string
 there are no new BPS states other
 than the usual ones corresponding to operators $Tr[Z^J]$.

 Notice that the classical string formula \nref{energy} is missing
 a 1 in the square root as compared to \nref{disprel}. This is no
 contradiction since we were taking $p$ fixed and $\lambda$ large
 when we did the classical computation. This 1 should appear after we
 quantize the system. In fact, for small $p$
 and $\lambda $ large, we can make a plane wave approximation and,
 after quantization, we recover the 1 \cite{metsaev,bmn}.
  But if we did not quantize we would not
 get the 1, even in the plane wave limit.  So we see that in the
 regime that the 1 is important we indeed recover it by doing the
 semiclassical quantization. This 1 is also implied by the
 supersymmetry algebra. The argument is identical to the one in
 \cite{beiserts} once we realize that the central charges are
 present and we  know the relation between the
 central charges and the momentum $p$, as in \nref{momcen}.
 Notice that the classical solutions we discussed above break the
 $SO(4)$ symmetry since they involve picking a point on $S^3
 \subset S^5$ where the straight string in figure \ref{circle}(a)
 is sitting. Upon collective coordinate quantization we expect that the string
 wavefunction will be constant on this $S^3$. In addition, we
 expect to have fermion zero modes. They originate from the fact
 that the magnon breaks half of the 16 supersymmetries that are
 left unbroken by the string ground state. Thus we expect 8
 fermion zero modes, which, after quantization, will give rise to
 $2^4 = 16 $ states, 8 fermions + 8 bosons. This argument is
 correct for fixed $p$ and large $\lambda$.
In fact, all these arguments are essentially the same as the ones
we would make for a string stretching between two D-branes. Notice
that all magnons look like stretched strings in the $S^5$
directions, as in figure \ref{circle}(a), including magnons
corresponding to insertions of $\partial_\mu Z$ which parameterize
elementary excitations along the $AdS_5$ directions. Of course,
here we are considering a single magnon. Configurations with many
magnons can have large excursions into the $AdS$ directions.

Since the stretched string solution in figure \ref{circle}(a) is
BPS, it is the minimum energy state for a given $p$.

 In fact, we can consider large $J$ string states around other
 1/2 BPS geometries, given by different droplet shapes as in
 \cite{llm}. In those cases, we will have BPS configurations
 corresponding to strings ending at different points on the
 boundary of the droplets and we have strings stretching between
 these points.  It would be nice to see if the resulting
 worldsheet model is integrable.

Note that the fact that the magnons have a large size (are
``giant'') at strong coupling is also present in the Hubbard model
description in \cite{hubbard} \footnote{ We thank M. Staudacher
for pointing this out to us. }.

Finally, let us point out that our discussion of the classical
string solutions focussed on an $R\times S^2$ subspace of the
geometry. Therefore, the same solutions will describe giant
magnons in the plane wave matrix model \cite{bmn,plefkamm} and
other related theories \cite{lm}. Similar solutions also exist in
$AdS_{2,3} \times S^{2,3}$ with RR fluxes (NS-fluxes would change
the equations already at the classical level).

 \subsection{ Spinning folded string}

\begin{figure}[htb]
\begin{center}
 \epsfxsize=1in\leavevmode\epsfbox{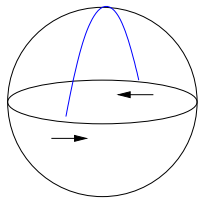}
\end{center}
\caption{ Spinning string configuration that corresponds to two
magnons with $p= \pi$.
 } \label{spinning}
\end{figure}

 In this subsection we apply the ideas discussed above to compute
 the energy of a spinning folded string considered in \cite{gkpspin}. This is a
 string that rotates in an $S^2$ inside $S^5$. For small angular
 momentum $J$  this is a string rotating around the north
 pole. Here we are interested in the limit of large $J$ where the
 ends of the rotating string
 approach the equator, see figure \ref{spinning}. In this limit
 the energy of the string becomes \cite{gkpspin}
\be \label{gkpres}
  E- J = 2 { \sqrt{\lambda } \over \pi }
  \ee

 This string corresponds to a superposition of two  ``magnons''
 each with maximum momentum, $p=\pi$.  Notice that the dispersion relation implies
 that such magnons are at rest, see \nref{velgauge}.
  They are equally spaced on the
 worldsheet. At large $J$ we can ignore
 the interaction between the ``magnons'' and compute the energy of
 the state as a superposition of two magnons.
 We see that the energy \nref{gkpres} agrees precisely with the
 energy of two magnons with $p=\pi$.

  Since the $\lambda$ dependence of the strings spinning in AdS in
  \cite{gkp}
    is
  somewhat similar, one might find an argument for that case too.
In fact, the solutions we are considering here, such as the one in
figure \ref{circle}(b) is the sphere analog of the solutions with
spikes considered in \cite{kruczenski}.

  \section{Semiclassical S-matrix}

 \subsection{General constraints on the S-matrix}

 In this section we consider the S-matrix for scattering of two
 magnons. On the gauge theory side this is the so called
 ``asymptotic S-matrix'' discussed in \cite{staudachers}. In the
 string theory side it is defined in a similar way: we take two
 magnons and scatter them. Then, we define the $S$ matrix for
 asymptotic states as we normally do in 1+1  dimensional field
 theories. Since the sigma model is integrable \cite{Mandal,bpr}, we
 expect to have factorized scattering. It was shown in
 \cite{frolovint} that integrability still persists in the lightcone
 gauge (this was shown ignoring the fermions). In fact we will
 later check explicitly that our magnons undergo classical dispersionless
 scattering.

\begin{figure}[htb]
\begin{center}
 \epsfxsize=5in\leavevmode\epsfbox{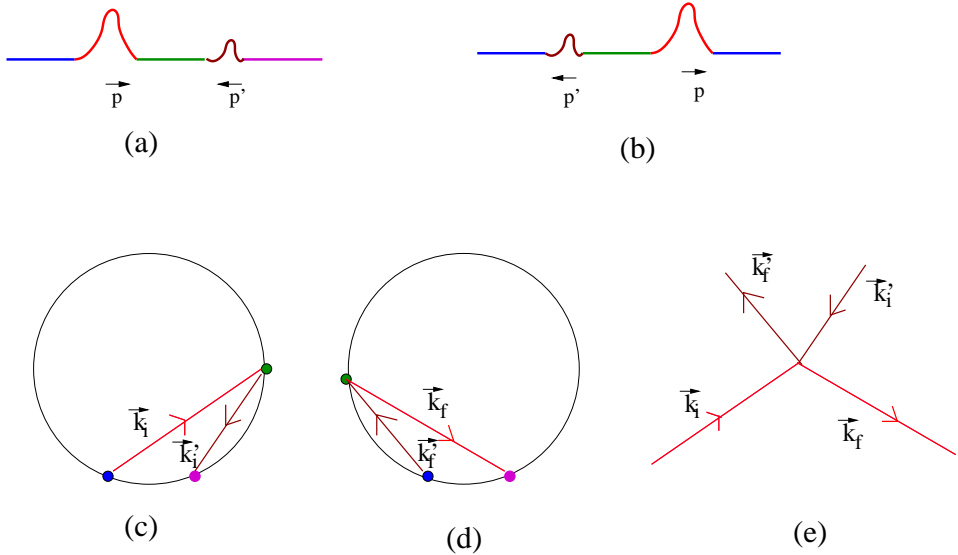}
\end{center}
\caption{Scattering of $2 \rightarrow 2$ magnons. (a) Worldsheet
picture of asymptotic initial state. (b) Worldsheet picture of
asymptotic final state. (c) Initial state in LLM coordinates. (d)
Final state in LLM coordinates. (e) Initial and final
configuration for the momenta $\vec k$ that are relevant for the
2+1 dimensional kinematics of the process.
 } \label{scattering}
\end{figure}

  As we mentioned above the supersymmetry algebra is the same in
  the gauge theory and the string theory. We have only shown here
  that the algebra is the same at the classical level on the
  string theory side. But it is very natural to think that after
  quantization we will still  have the same algebra.
  Thus, any constraint
  coming from this algebra is the same. An important
  constraint for the $S$ matrix
  was derived by Beisert in \cite{beiserts}.
   Each of the magnons
  can be in one of 16 states (8 bosons plus 8 fermions).
   So the scattering matrix is a
  $16^2 \times 16^2$ matrix. Beisert showed that this matrix is
  completely fixed by the symmetry up to an overall phase (and
  some phases that can be absorbed in field redefinitions). Schematically
  $S = \hat S_{ij} S_0$ where $\hat S_{ij}$ is a known matrix and $S_0$ is an
  unknown phase.  The
  same result holds then for the string theory magnons. In fact,
  it was conjectured in \cite{stringbethe,staudachers,nbms}
  that the two S-matrices differ by
  a phase. Here we are pointing out that this structure is a
  consequence of the symmetries on the two sides.
  The fact that the whole $S$ matrix is determined up to a
  single function is analogous to the statement that the four particle
  scattering amplitude in ${\cal N}=4$ SYM is fixed up to a
  scalar function of the kinematic invariants. The reason is that
  two massless particles with 16 states each
  give a single  massive, non-BPS,
  representation with $2^8 = 16^2$ states.

A two magnon scattering process has a kinematics that is shown in
figure \ref{scattering}.  Notice that we can literally think of
the straight strings as determining the initial and final momentum
vectors of the scattering process as in figure
\ref{scattering}(e). The orientation of these vectors is
important. The constraints on the matrix structure of the $S$
matrix are exactly the same as the constraints that a four
particle scattering amplitude in a relativistic 2+1 dimensional
field theory with the same superalgebra would have. These
constraints are easy to derive in the center of mass frame. And we
could then boost to a general frame. Notice that from the 2+1
dimensional point of view fermions have spin, and thus their
states acquire extra phases under rotation. In other words, when
we label a state by saying what its momentum $p$ is, we are just
giving the magnitude of $\vec k$, but not its orientation. The
orientation of $\vec k$ depends on the other magnons. For example,
in the scattering process of figure \ref{scattering}(a,b) the
initial and final states have the same momenta $p, p'$, but the
initial vectors ${\vec k}_i,~{\vec k'}_i$ have different
orientation than the final vectors ${\vec k}_f,~{\vec k'}_f$. When
we consider a sequence of scattering processes, one after the
other, it is important to keep track of the orientation of $\vec
k$. In other words, under an overall rotation the $S$ matrix is
not invariant, it picks up some phases due to the fermion spins.
In \cite{beiserts} these phases are taken into account by extra
insertions of the field $Z$ which makes the chain ``dynamic''.

Note that the constraints on the matrix structure of the
scattering amplitude are applicable in a more general context to
any droplet configuration of \cite{llm}.
 For example, it constrains the scattering amplitude for elementary excitations
 in other theories with the same superalgebra. Examples are the massive M2 brane theory
 \cite{popewarner} or the theories considered in
 \cite{lm,seiberg}.

  Note that the existence
  of closed subsectors is a property of factorized scattering (integrability)
  and the matrix structure of the
  $\hat S$ matrix,  but does not depend on the precise nature of
  the overall phase.
   Thus closed subsectors exist on both sides. This argument shows
   this only in the large $J$ limit where the magnons are well separated and
   we can use the asymptotic $S$ matrix\footnote{Note that this
   is not obviously in contradiction with the arguments against closed subsectors on the
   string theory side that were made in \cite{minahansec}, which
   considered
     finite $J$ configurations.}.

  We expect that the overall phase, $S_0$, will interpolate between the weak and
  strong coupling results.  The full
  interpolating function has not yet been determined\footnote{There
  is of course (the very unlikely possibility) that the two phases
  are different and that $AdS/CFT$ is wrong. }.

  In this section we will compute in a direct, and rather
  straightforward way, the semiclassical S-matrix for the
  scattering of string theory magnons. It turns out that the
  result will agree with the one derived in \cite{stringbethe}
  through more indirect methods.

 Notice that at large 't Hooft coupling and fixed momentum, the
 approximate expression \nref{larget} amounts to a relativistic
 approximation to the non-relativistic formula \nref{disprel}.
 Similarly, in this limit, the matrix prefactor $\hat S$ becomes
 that of a relativistic theory and it is a bit simpler.

  Notice that the theory in light cone gauge is essentially
  massive so that we can define scattering processes in a rather
  sharp fashion, in contrast with the full covariant sigma model
  which is conformal, a fact that complicates the scattering
  picture. Nevertheless,  starting from the conformal sigma model
  can be a useful way to proceed \cite{mannpolch}.

\subsection{Scattering phase at large $\lambda$ and the sine-Gordon
connection}

In the semiclassical limit where $\lambda$ is large and $p$ is
kept fixed   the leading contribution to the $S$ matrix
 comes
from the phase $\delta$ in $S_0 = e^{i \delta}$, which goes as
$\delta \sim \sqrt{\lambda} f(p,p')$. For fixed momenta, this
phase dominates over the terms that come from the matrix structure
$\hat S$ in the scattering matrix. In this section, we compute
this phase, ignoring the matrix prefactor $\hat S$ in the
S-matrix.

  In the semiclassical approximation the phase shift can be
  computed by calculating  the time delay that is accumulated when
  two magnons scatter. The computation is very similar to the
  computation of the semiclassical phase for the scattering of two
  sine Gordon solitons, as computed in \cite{jackiw}.
  In fact, the computation is almost {\it identical} because  the
  magnons we discuss are in direct correspondence with sine Gordon
  solitons. This uses the relation between classical sine Gordon
  theory and classical string theory on $R \times S^2 $
  \cite{pohlmeyer,mikhailov}\footnote{ As explained in
  \cite{mikhailovpoisson} the two theories have different poisson structures so
  that their quantum versions are different.}. It is probably
  also possible to obtain these results from \cite{kkmz}, but we
  found it easier to do it using the correspondence to the sine
  Gordon theory.
   The map between a classical string theory on $R \times S^2$ and the
   sine Gordon model goes as follows.
  We consider the string action in conformal gauge and we
  set $t = \tau$. Then the Virasoro constraints become
  \be
   1 = \dot {\bf n}^2 + {{\bf n}'}^2 ~,~~~~~\dot {\bf n} . {\bf n} '
   =0
   \ee
   where ${\bf n}^2 =1$ parameterizes the $S^2$.
   The equations of motion follow from these constraints. The sine
   Gordon field is defined via
   \be
   \cos 2 \phi =\dot {\bf n}^2 - {{\bf n}'}^2 \label{sgdefi}
   \ee
   For the ``magnon'' solution we had above we find that $\phi$ is
  the sine Gordon soliton
   \be
   \tan { \phi \over 2 } = \exp\left[{ \cos{ p\over 2} t - x \over
   \sin\p2 }\right] = e^{- \gamma ( x - v t )} ~,~~~~~~~~~ v = \cos \p2
   ~,~~~~~~~~ \gamma^{-2} = 1 - v^2
   \ee
   Notice that the energy of the sine Gordon soliton is inversely
   proportional to the string theory energy of the excitation
   \nref{energy}
   \be \label{rapiditymom}
    E_{s.g.} = { \gamma} = \cosh \hat \theta ~,~~~~~~~~~\epsilon_{magnon} =
     { \sqrt{ \lambda} \over \pi} \, { 1 \over
    \gamma} ~,~~~~~~~~~~~~~\cosh \htheta = { 1 \over \sin \p2 }
    \ee
    where we measure the sine Gordon energy relative to the energy
    of a soliton at rest and we introduced the sine Gordon
    rapidity $\htheta$.
    Note that a boost on the sine Gordon side translates into a
    non-obvious {\it  classical}  symmetry on the $R\times S^2$ side.
 Do not confuse this approximate boost symmetry of the sine Gordon
  theory with the boosts that appeared in our discussion of the
  supersymmetry algebra. Neither of them is a true symmetry of the
  problem, but they are not the same!.

   We  now consider a soliton anti-soliton pair and we
   compute the time delay for their scattering as in \cite{jackiw}.
 (If we use a soliton-soliton pair we obtain the same
 classical
 answer\footnote{ In fact, for a given $p$ we have a family of magnons given by a choice
 of a point on $S^3$ which is telling us how the string is embedded in $S^5$.
 In the quantum problem this zero mode is quantized and the wavefunction will be spread on
 $S^3$.
 In the classical theory we expect to find the same time delay for scattering of two magnons
 associated to two arbitrary points on $S^3$.}).
    Since the $x$ and
   $t$ coordinates are the same in the two theories,
   this time delay is precisely the
   same  for the string theory magnons and for the sine Gordon
   solitons. The Sine Gordon  scattering solution  in the center of mass
   frame is
   \be \tan {\phi \over 2 } = { 1 \over v} { \sinh \gamma v t
   \over \cosh \gamma x } \label{ssbar}
   \ee
   The fact that the sine Gordon scattering is dispersionless
   implies that the scattering of magnons is also dispersionless
   in the classical limit (of course we also expect it  to
   be dispersionless in the quantum theory).

   The time delay is
   \be
   \Delta T_{CM} =   { 2 \over \gamma v } \log v
   \ee
   We now boost the configuration \nref{ssbar} to a frame
   where we have a soliton moving with velocity $v_1$ and an
   anti-soliton
   with velocity $v_2$, with $v_1>v_2$.
   Then the time delay that particle 1 experiences as it goes
   through particle 2 is
   \be
   \Delta T_{12} = { 2 \over \gamma_1 v_1 } \log v_{cm}
   \ee
   where $v$ is the velocity in the center of mass frame
   \be
   2 \log v_{cm} = 2 \log  \tanh \left[ { \hat \theta_1 - \hat \theta_2 \over 2}
   \right] =    \log \left[ { 1 - \cos{ p_1- p_2 \over 2} \over
  1 - \cos { p_1 + p_2 \over 2} } \right] ~,~~~~~{\rm
  for}~~~p_1,~p_2 >0
   \ee
   We can now compute the phase shift from the formula
   \be
    { \partial  \delta_{12}(\epsilon_1,\epsilon_2) \over \partial
    \epsilon_1} =  \Delta T_{12}
    \ee
    We obtain
   \be \label{ourphase}
   \delta = { \sqrt{\lambda} \over \pi } \left\{
    \left[  - \cos{ p_1 \over 2} + \cos { p_2 \over 2 } \right]
  \log \left[ { 1 - \cos{ p_1- p_2 \over 2} \over
  1 - \cos { p_1 + p_2 \over 2} } \right] \right\}
  -  p_1   { \sqrt{\lambda} \over \pi } \sin {p_2 \over 2}
   \ee
   Note that, even though the time delay is identical to the sine
   Gordon one, the phase shift is different, due to the different
   expression for the energy \nref{rapiditymom}.
    This implies, in particular, that
   the phase shift is not invariant under sine Gordon boosts.
The first term in this expression agrees precisely with the large
$\lambda$ limit of the phase in \cite{stringbethe}\footnote{ The
phase in \cite{stringbethe} contains further terms in a
$1/\sqrt{\lambda}$ expansion which we are not checking here.}.
 The
second term in \nref{ourphase} looks a bit funny. However, we need
to recall that the definition of this S-matrix is a bit ambiguous.
This ambiguity is easy to see in the string theory side and was
noticed before. For example \cite{swanson} and \cite{frolovfer}
obtained different S-matrices for the scattering of magnons at low
momentum (near plane wave limit). The difference is due to a
different choice of gauge which translates into a different choice
of worldsheet $x$ variable. In \cite{swanson} the $x$ variable was
defined in such a way that the density of $J$ is constant. In
\cite{frolovfer} it was defined so that the density of $E+J$ is
constant. In our case we have defined it in such a way that the
density of $E$ is constant, since we have set $\dot t =1$ in
conformal gauge. All these choices give the same definition for
the $x$ variable when we  consider the string ground state. The
difference lies in the different length in $x$ that is assigned to
the magnons, which have $E-J\not =0$. Thus the $S$ matrix computed
in different gauges will differ simply by terms of the form $e^{i
p_i f(p_j)}$ where $f(p_j)$ is the difference in the length of the
magnon on the two gauges. Of course the Bethe equations are the
same in both cases since the total length of the chain is also
different and this cancels the extra terms in the $S$ matrix. The
position variable that is usually chosen on the gauge theory side
assigns a length 1 to the impurity. At large $\lambda$ we can
ignore 1 relative to $\lambda$ and say that the length of the
impurity  is essentially zero. Thus we can say that the gauge
theory computation is using coordinates where the density of $J$
is constant.  Using the relation between the gauge theory spatial
coordinate $l$ and our worldsheet coordinate $x$ \nref{xandl}
(which is valid in the region where $E-J=0$) we get that the
interval between  of two points separated by a magnon are related
by
 \be \label{landxen}
 \Delta l =    \int d x { d J \over d
 x }  =    \int d x {d E \over d
 x} - ( { dE \over d x} - { d J \over d x} )  =
{ 2 \pi \over \sqrt{\lambda} }  \Delta x - \epsilon   \ee where
$\Delta l$ is the interval in the conventions of
\cite{stringbethe} and $\Delta x$ is the interval in our
conventions.
 So we see that in our gauge the magnon will have an extra length
 of order $\epsilon(p)$. Thus $S_{string-Bethe} = S_{ours}
 e^{i p_1 \epsilon_2 } $, where $S_{string-Bethe}$ is the $S$ matrix in the conventions
 used in \cite{stringbethe}.
This cancels the  last term in
 \nref{ourphase}.
 In summary, after expressing the result in conventions adapted
 to the gauge theory computation we find that for
 $sign (\sin {p_1 \over 2}) >0$ and $sign(\sin { p_2 \over 2} )> 0$ we get
 \be \label{ourfinal}
 \delta(p_1,p_2) = - {\sqrt{\lambda } \over \pi } ( \cos{p_1\over 2} - \cos {
 p_2
 \over 2} ) \log \left[ { \sin^2 { p_1-p_2 \over 4} \over \sin^2 {
 p_1
 + p_2 \over 4 } } \right]
 \ee
The cases where $p<0$ can be recovered by shifting $p$ by a period
so that $\sin { 2 \pi + p \over 2} >0$.
 The function \nref{ourfinal} should be trusted when $\sin {p_i
 \over 2 } >0$ and it should be defined to be periodic with period
 $ 2 \pi$ outside this range.
   Note that this function goes to zero when
 $p_1 \to 0$ with $p_2$ fixed. When $p$ is small we need to quantize
 the system. We can check that, after quantization, the $S$ matrix
 is still trivial for small $p_1$ and fixed $p_2$. This can be done
 by expanding in small fluctuations around our soliton background.
 We find that the small excitations propagate freely through the
 soliton.

The leading answer  \nref{ourfinal} vanishes at small $p$.    In
fact, at small $p$ it is important to properly quantize the system
and the result  depends on the polarizations of the states, see
\cite{swanson}. For example, in the $SU(1|1)$ sector
\cite{swanson} found (see also \cite{frolovfer})
 \be
 S_{string} = -1 +  i {1 \over 2} \left(   p_1 \sqrt{ 1 + {\lambda \over 4 \pi^2
 } p_2^2 } - p_2 \sqrt{ 1 + {\lambda \over 4 \pi^2
 } p_1^2 }  - p_1 + p_2 \right)
 \ee

 Corrections to the leading phase \nref{ourfinal} were computed
in \cite{lopez} and some checks were made in
\cite{checklopez,frolovcheck}.

On the gauge theory side the phase is known up to three loop
orders in $\lambda$ \cite{3loops}. Of course, finding the full
interpolating function is an outstanding challenge\footnote{An all
loop guess was made in \cite{bds} (see also \cite{hubbard}),  but
this guess appears
 to be in conflict with the strong coupling
results obtained via $AdS/CFT$.}.

\begin{figure}[htb]
\begin{center}
\epsfxsize=6in\leavevmode\epsfbox{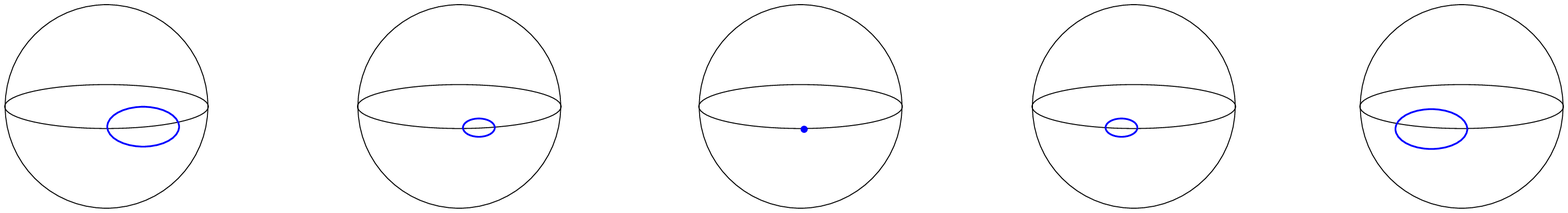}
\end{center}
\caption{Evolution of the soliton anti-soliton scattering state.
Time increases to the right. We have chosen a rotating frame on
the sphere where the point with infinite $J$ is stationary.
 At $t=0$ the string is all concentrated at a point,
the point that carries infinite $J$. The state looks
asymptotically as two free magnons on opposite hemispheres and
with the same endpoints.} \label{scattersa}
\end{figure}

Finally, to complete the discussion of scattering states, we
comment on the spacetime picture of the scattering process. In the
classical theory, besides specifying $p$, we can also specify a
point on $S^3$ for each of the two magnons that are scattering off
each other. We do not know the general solution. The sine Gordon
analysis we did above applies only if the point on $S^3$ is the
same for the two magnons or are antipodal for the two magnons. In
the first case we have a soliton-soliton scattering in the
sine-Gordon model and in the second we have a soliton-anti-soliton
scattering. Both give the same classical time delay.
 The
soliton anti-soliton scattering with $p_1 = - p_2$ looks initially
like loop of string made of two magnons. One of the endpoints has
infinite $J$ and the other has finite $J$.  The point in the
front, which initially carries a finite amount of $J$, looses all
its $J$ and it moves to the left. The loop becomes a point and
then the loops get formed again but with the finite $J$ point to
the left, behind the point that carries infinite $J$. See figure
\ref{scattersa}. The soliton soliton scattering is represented by
a doubly folded string that looks initially like a two magnon
state. As time evolves the point in the front, which carries
finite $J$,  detaches from the equator and moves back of the other
endpoint which carries infinite $J$. The final picture is, again,
equivalent to the original one with front and back points
exchanged. We see that in both cases asymptotic states are well
defined and look like individual magnons.

  \subsection{Bound states}

One interesting property of the sine Gordon theory is that it
displays an interesting set of localized states which can be
viewed as soliton-antisoliton bound states. The corresponding
classical solutions are given by performing an analytic
continuation $v \to i a $ in \nref{ssbar}
 \be \label{sgbound}
 \tan {\phi \over 2} = { 1 \over a } { \sin \gamma_a a t \over
 \cosh \gamma_a x } ~,~~~~~~~\gamma_a^{-2} = 1 + a^2
  \ee
  In the semiclassical
 approximation to the sine-Gordon theory we can quantize these
 modes \cite{jackiw,dashen}. These particles then appear as poles
 in the $S$ matrix, \cite{zamolodchikov}.

In this section we will use these sine Gordon solutions
\nref{sgbound} in order to produce solutions representing bound
states of magnons. We will then quantize them semiclassically.

 We can start with the solution \nref{sgbound} and we boost it with a
 boost parameter $\gamma$. Then the soliton and anti-soliton components have
 rapidities
 \be \label{rapid}
 \hat  \theta_{1,2} = \hat \theta \pm i \hat \theta_a ~,~~~~~~~
 v = \tanh \htheta ~,~~~~~~~~ a = \tan \htheta_a
 \ee
 where $v$ characterizes the velocity of the bound state and $a$
 is the parameter $a$ appearing in the center of mass frame
 solution
 \nref{sgbound}. In the sine Gordon theory the energy is given by
 the sum of the energies of a soliton anti-soliton pair with these
 rapidities \nref{rapid}
  \be
  E^{bound}_{sg} = E(\hat \theta + i \hat \theta_a) + E(\hat \theta - i \hat \theta_a)
  = \cosh (\hat \theta + i \hat \theta_a) + \cosh(\htheta - i \htheta_a) =
  2 \cosh \htheta \cos \htheta_a
  \ee
  We can understand this formula as resulting from analytic
  continuation of the formula that gives a similar additive result for
  scattering states. Recall that \nref{sgbound} is a configuration
  obtained from a scattering solution \nref{ssbar} by analytically
  continuing the rapidities.

 We now want to understand what these states correspond to in our
 system. From the string theory point of view we can label the
 states in terms of the momentum $p$ or in terms of the rapidity
 $\hat \theta$ related by
 \nref{rapiditymom}.
 The localized solutions \nref{sgbound} should correspond to
 localized solutions on the worldsheet that come from analytically
 continuing the parameters of scattering solutions. Thus we expect
 that, also in the string theory,  the energy will be a sum of the
 energies of its components
 \be \label{bounden}
 \epsilon^{bound} = \epsilon (  \htheta_1) + \epsilon( \htheta_2)
 = { \sqrt{ \lambda } \over \pi } \left[ { 1 \over \cosh \htheta_1 } +
 { 1 \over \cosh \htheta_2 } \right]
 \ee
 This expression for the energy is important for what we will do
 later. We have checked directly that starting with a boosted
 version of \nref{sgbound}, inverting the equation \nref{sgdefi},
 and computing the energies we get \nref{bounden}.
 See appendix B for more details.
 From the rapidities we can also define the momenta of each of the
 particles through the relation \nref{rapiditymom}. These momenta
 are complex
 \be \label{raprel}
  p_{1,2} = p  \pm i q ~,~~~~~~~~ { 1 \over \cosh \htheta_i} =
  \sin { p_i \over 2}
  \ee
We also see that the total momentum of the state is $P = 2 p$.

In order to semiclassically quantize these states we need to find
the period of oscillation. This is the time it takes for the
solution to look the same up to an overall translation.   We find
 \be
 \label{period}
 T = 2 \pi { \gamma \over \gamma_a a } = 2 \pi  { \cosh \htheta \over
 \sin \htheta_a } = { 2 \pi } { 1 \over \tanh { q \over 2} }
~,~~~~~~~~{\rm note~that} ~~~\tan \p2 = { \gamma_a \over v \gamma}
\ee where we used \nref{raprel}. We now define the action variable
$n$ which in the quantum theory should be an integer, through the
equation \be
 dn = { T \over 2 \pi } d \epsilon|_{p} =  { \sqrt{\lambda } \over
 \pi} \sin \p2 \cosh { q \over 2} dq ~~~~{\rm or}~~~ n = {2  \sqrt{\lambda } \over
 \pi} \sin \p2 \sinh { q \over 2}
 \ee
where we keep the momentum fixed, which is another conserved
quantity. We obtain
 \be
 \epsilon = { \sqrt{\lambda } \over
 \pi} ( \sin {p_1 \over 2} + \sin {p_2 \over 2} ) =
 { 2 \sqrt{\lambda } \over
 \pi} \sin \p2 \cosh { q \over 2} =
  \sqrt{ n^2 + { 4 \lambda \over \pi^2 } \sin^2 \p2    }
 \ee
 where we should trust this formula only for large $n$. So, in the
 regime where we can trust it, these states have more energy than
 two magnons each with momentum $p$. We  do not see any sign of a
 breakdown in our analysis for very large $n$. So
 for large but finite $\lambda$ there is an infinite number of
 bound states. It looks like these bound states
should belong to general massive representations of $SU(2|2)^2$
since those are the ones that generally correspond to a two magnon
configuration.

 Note that the bound states carry momentum $P = 2p$ and that as
 $p$ varies from zero to $\pi $, $P$ varies from 0 to
 $ 2 \pi$ and the two endpoints of the
 string move over the whole equator of the $S^2$.
 But the configuration is not periodic in $P$. Namely, for small
 $P$ we have a bound state of two magnons with small $p$, while
 for $P \sim 2 \pi$ we have a bound state of two large magnons
 with $p \sim \pi $.

\begin{figure}[htb]
\begin{center}
\epsfxsize=4in\leavevmode\epsfbox{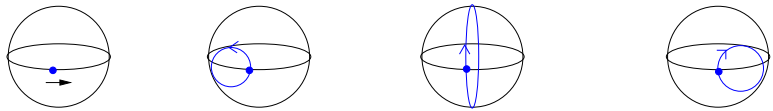}
\end{center}
\caption{Evolution of the bound state with $p=\pi$. Time increases
to the right. We have chosen a rotating frame on the sphere where
the point with infinite $J$ is stationary.
 At $t=0$ the string is all concentrated at a point,
the point that carries infinite $J$. As time progresses a loop
forms, then this loop sweeps the whole sphere and becomes a point
again. The motion repeats itself again except that the loop has
the opposite orientation. Thus in each period we performs two
sweeps of the sphere.  } \label{boundfi}
\end{figure}

\begin{figure}[htb]
\begin{center}
\epsfxsize=2in\leavevmode\epsfbox{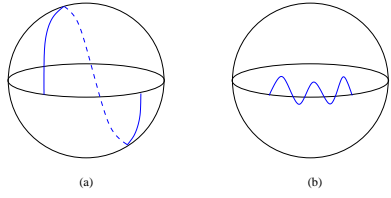}
\end{center}
\caption{Snapshots of other string bound states.  (a) Large p
bound state. (b) Small p bound state.} \label{boundsta}
\end{figure}

The spacetime picture of these solutions varies considerably
depending on parameters $p$ and $q$. The easiest case to analyze
is the solution corresponding to the breather at rest. This
corresponds to  choosing maximal $p=\pi$. These are strings with
one fixed point ($\varphi-t=\textrm{const}$) which   sweep  the
entire sphere as they evolve in time, see figure \ref{boundfi}. At
quarter the period  they look like two magnons of maximal $p$. The
value of $q$ controls the period of the sweep.    Because
$P=2p=2\pi$ for this case there is no distance between the
endpoints of the string. As we decrease $p$ a gap opens up while
the strings still sweep the sphere, see figure \ref{boundsta}(a).
At $p=\frac{\pi}{2}$ the gap is maximal and the solutions change
character:they do not sweep the sphere any longer. For small $p$
the solution is bounded to a small region of the sphere, see
figure \ref{boundsta}(b). $q$ still controls the period. In
appendix B, we discuss the relevant variables and calculate the
energies of these solutions. It would be nice to find more
explicit expressions for the solutions.

 We can now
compare these results to the bound states at weak coupling. We
focus on the $SU(2)$ sector for simplicity and to lowest order in
the 't Hooft coupling. In that case we simply have the XXX spin
chain \cite{minzar}. It is then easy to show by looking at the
poles of the $S$ matrix that there is single bound state of two
magnons. Writing the momenta as $p_{1,2} = p \pm i q$ one can
check that the bound state has
 \be  \label{invsize}
  q = \log (\cos p)
 \ee
Using the dispersion relations for magnons we can check that the
energy is
 \be
 \epsilon = \epsilon(p_1) + \epsilon(p_2) = \sum_{i=1}^2 1 + { \lambda   \over 2 \pi^2}
 \sin^2 { p_i \over 2}  = 2 + { \lambda \over 2 \pi^2
 } 2 \cos^2 \p2 \sin^2 \p2
 \ee
 In this case we see that the energy is smaller than the sum of
 the energies of two magnons with momentum $p$. We also see that
 the size of the bound state goes to infinity, and the binding
 energy disappears, when $q\to 0$. In other sectors, such as the
 $SU(1|1)$ or $SL(2)$ sectors there are no bound states.

 We see that the number of bound states is very different at weak
 and strong coupling. Presumably, as we increase the coupling new
 bound states appear. In the ordinary sine Gordon theory bound
 states disappear as we increase the coupling, we start with a
 large number at weak coupling and at strong coupling there are
 none, we are left just with the solitons. A new feature of our
 case is that the number of bound states is  infinite at large $\lambda $.
  So at
 some value of the coupling we should be getting an infinite
 number of new bound states. The situation is somewhat similar to
 the spectrum of excitations of a quark anti-quark pair in
 \cite{thorn}.
It turns out that states with large $n$ have large size, so that
we need large $J$ to describe them accurately enough.

\section{Discussion}

In this article we have introduced a limit which allows us to
isolate quantum effects from finite volume effects in the gauge
theory/spin chain/string duality. In this limit, the symmetry
algebra is larger than what is naively expected. This algebra is a
curious type of 2+1 superpoincare algebra, without the lorentz
generators, which are not a symmetry. The algebra is the same on
both sides.
 In this infinite $J$
limit the fundamental excitation is the ``magnon'' which is now
identified on both sides. The basic observable   is the scattering
amplitude of many magnons. Integrability should imply that these
magnons obey factorized scattering so that all amplitudes are
determined by the scattering matrix of fundamental magnons. The
matrix structure of this S-matrix is determined by the symmetry at
all values of the coupling. So the whole problem boils down to
computing the scattering phase \cite{beiserts}. This phase is a
function of the two momenta of the magnons and the 't Hooft
coupling. At weak coupling it was determined up to three loops
\cite{3loops}. At strong coupling we have the leading order
result, computed directly here and indirectly in
\cite{stringbethe} (see also \cite{swanson}). The one loop sigma
model correction to the S-matrix was computed using similar
methods in \cite{lopez}. As in other integrable models, it is very
likely that a clever use of crossing symmetry plus a clever choice
of variables would enable the computation of the phase at all
values of the coupling. Recently, a crossing symmetry equation was
written by Janik \cite{janik}. The kinematics of this problem is a
bit different than that of ordinary relativistic 1+1 dimensional
theories. In fact, the kinematic configuration has  a double
periodicity \cite{janik}. This is most clear when we define a new
variable $\theta_p$ as \be
  |\vec k|^2 = { \lambda \over \pi^2 } \sin^2 \p2 = \sinh^2
 \theta_p ~,~~~~~~\epsilon = \cosh \theta_p
 \ee
  We have a periodicity in $ \theta_p
\to \theta_p +2 \pi i$ and in $p \to p + 2 \pi $. Crossing is
related to the change $\theta_p \to \theta_p +  i \pi $. The full
amplitude does not need to be periodic in these variables since
there can be branch cuts. Of course, one would like to choose a
uniformizing parameter that is such that the amplitude becomes
meromorphic. A proposal for one such parameter was made in
\cite{janik}. Perhaps the 2+1 dimensional point of view might be
useful for shedding light on the choice of  variables to describe
the scattering process. An additional complication is that the
S-matrix appears to depend on the  two momenta, rather than a
single variable (the center of mass momentum). On the string
theory side it is very reasonable to expect some type of crossing
symmetry due to the form of the dispersion relation, which is
quadratic in the energy. Indeed, crossing is a property of the
first two orders in the coupling constant expansion away from
strong coupling \cite{frolovcheck}. At weak coupling crossing
symmetry is not manifest because the weak coupling expansion
amounts to a non-relativistic expansion, but it should presumably
be a symmetry once the full answer is found.

The bound states that we have encountered at strong coupling
should appear as poles in the exact S-matrix\footnote{These are
not present in the semiclassical result \nref{ourfinal}.}.
Presumably, as we increase the 't Hooft coupling we will have, at
some point, an infinite number of poles appearing. In the sine
Gordon model the number of bound states changes as we change the
value of the coupling \cite{zamolodchikov}. But we always have a
finite number.

We should finally mention that perhaps it might end up being most
convenient to think of the problem in such a way that the magnon
will be composed of some more elementary excitations, as is the
case  in \cite{hubbard}\footnote{ The equivalence \cite{hubbard}
between the Hubbard model and the gauge theory holds up to 3 loop
orders. Beyond 3 loops it agrees with the all loop guess in
\cite{bds}, but the guess in \cite{bds} seems to be in conflict
with string theory at large $\lambda$.}. On the other hand, we do
not expect these more elementary excitations to independently
propagate along the chain. For this reason it is not obvious how
to match to something on the string theory side.

{\bf Acknowledgments }

We would like to thank N. Beisert, S. Frolov, K. Intrilligator, J.
Plefka, N. Seiberg, M. Staudacher and I. Swanson for useful
comments and discussion.

This work was supported in part by DOE grant \#DE-FG02-90ER40542.

\section{Appendix A: The supersymmetry algebra}

We start with a single $SU(2|2)$ subgroup first. This algebra has
two $SU(2)$ generators and a non-compact generator $k^0 \equiv
\epsilon = E-J$. The superalgebra is
 \bea
 \{ {Q'}^{bs},Q^{ar} \} &=& \epsilon^{ba} \epsilon^{sr} k^0 + 2[
 \epsilon^{ba} J^{sr} - \epsilon^{sr} J^{ba} ]
\\ \label{naivezero}
 \{ {Q}^{bs},Q^{ar} \} &=& =0 ~,~~~~~~~~~~
\{ {Q'}^{bs},{Q'}^{ar} \}  =0
 \eea
 (with $\epsilon^{+-}=1$) where we denote by $a,b$ the first $SU(2)$ indices
 and by $rs$ the second $SU(2)$ indices. We also have a reality condition
 $(Q^{ar})^\dagger = \epsilon_{ab} \epsilon_{rs} {Q'}^{b s} $.
 The central extensions considered in \cite{beiserts} involve two
 other central generators $k^1,k^2$ appearing on the right hand
 side of \nref{naivezero}, we will arbitrarily choose the
 normalization of these generators in order to simplify the
 algebra.
 In order to write the resulting algebra it is convenient to
 put together the two generators as
 \be
 q^{\alpha a r} = (Q^{a r}, {Q'}^{ar} ) ~,~{\rm or} ~~~~~q^{+ a r} = Q^{a r}
 ~,~~~~~~~~ q^{- a r} = {Q'}^{a r}\ee
where $\alpha, \beta$ will be $SL(2,R) = SO(2,1)$ indices. We
introduce the gamma matrices
 \be \label{gamdef}
  (\gamma_0 )_\alpha^{~ \beta} = i \sigma^3   ~,~~~~~(\gamma_1 )_\alpha^{~  \beta}
  =
  \sigma^1 ~,~~~~~(\gamma_2 )_\alpha^{~  \beta}
  =
  \sigma^2
 ~,~~~~~~~(\gamma_\mu)^{\alpha \beta }= \epsilon^{\alpha \delta}
 ( \gamma_\mu )_\delta^{~ \beta}
  \ee
 The full anti commutators will now have the form
 \be
 \{ q^{\alpha a r} , q^{\beta b s } \} = i \epsilon^{ba} \epsilon^{sr}
   ( \gamma_\mu )^{\alpha \beta} k^\mu  -
 2 \epsilon^{\alpha \beta}  [
 \epsilon^{ba} J^{sr} - \epsilon^{sr} J^{ba} ]
 \ee

The smallest representation of this algebra contains a bosonic
doublet and a fermionic doublet transforming as $(1,2)_b +
(2,1)_f$ under $SU(2)\times SU(2)$. If we think of these as
particles in three dimensions, then we also need to specify the
2+1 spin of the excitation. It is zero for the bosons and $+\half$
for the fermions. Let us call them $\phi^r$ and $\psi^a$. Notice
that this representation breaks parity in three dimensions. Once
we combine this with a second representation of the second
$SU(2|2)$ factor and the central extensions we obtain the eight
transverse bosons and fermions. We then have excitations $\phi^r
\tilde \phi^{r'}$ which are the bosons in the four  transverse
directions in the sphere. They have zero spin, which translates
into the fact that they have zero $J$ charge. This is as expected
for insertions of impurities of the type $X^i$, $i=1,2,3,4$
corresponding to the SO(6) scalars which have zero charge under
$J$.  We can also form the states $\psi^a \tilde \psi^{a'}$. These
have spin one from the 2+1 dimensional point of view. These states
are related to impurities of the form $\partial_i Z$, which have
$J=1$. States with a boson and a fermion, such as $\phi^r \tilde
\psi^{a'}$ or $\psi^a {\tilde \phi}^{r'}$ correspond to fermionic
impurities, which have $J=\half $. Notice that the spectrum is not
parity invariant, we lack particles with negative spin. This is
expected since parity in the $x_1,x_2$ plane of the   coordinates
in \cite{llm} is not a symmetry.

\section{Appendix B: Details about the analysis of bound states}

We choose coordinates for the two sphere so that
 \be
 ds^2 = d\theta^2 + \sin^2 \theta d \varphi^2
 \ee
 Given a solution of the string theory on $R \times S^2$ in
 conformal gauge, and with $t=\tau$, we define the sine Gordon
 field as
\be \label{sinegordde}
  \sin^2 \phi = {\theta'}^2 + \sin^2 \theta {\varphi '}^2
 \ee
We now want to invert this relation.
 From the Virasoro constraints we can solve for $\varphi'$ and
 $\dot \varphi$ in terms of $\theta$. We obtain
  \bea \notag
 \sin^2 \theta {\varphi'}^2 &=& \half \left[
 1 - \dot \theta^2 - {\theta'} ^2 - \sqrt{ (1- \dot \theta^2 +
 {\theta'}^2 )^2 - 4 {\theta'}^2 } \right]
 \\
 \sin^2 \theta {\dot \varphi}^2 &=& \half \left[
 1 - \dot \theta^2 - {\theta'} ^2 + \sqrt{ (1- \dot \theta^2 +
 {\theta'}^2 )^2 - 4 {\theta'}^2 } \right] \label{varphidef}
 \eea
 Inserting this into \nref{sinegordde} we get
 \be \label{sinegordnde}
  \sin^2 \phi = {\theta'}^2 + \sin^2 \theta {\varphi '}^2 =
\half \left[
 1 - \dot \theta^2 + {\theta'} ^2 - \sqrt{ (1- \dot \theta^2 +
 {\theta'}^2 )^2 - 4 {\theta'}^2 } \right]
 \ee

 Taking the time derivative of this equation we find something
 involving $\dot \phi$ in the left hand side and in the right hand
 side we will get terms involving $\ddot \theta$ and ${\dot
 \theta}'$. The terms involving $\ddot \theta$ can be eliminated
 by using the equation of motion so that we only have single time
 derivatives. Thus we will find an equation of the form
 \be
  \dot \phi = f( \theta, \dot \theta, \theta' , {\dot \theta}' )
  \ee
 This together with \nref{sinegordnde} determine $\theta$ and $\dot
 \theta$ at a particular time in terms of $\phi$ and $\dot \phi$ at that time.
  Note that to find $\theta, ~\dot \theta$  we will need
 to solve a differential equation in $x$.
 Once we determine $\theta$ and $\dot \theta$ at a particular
 time, we can use the equation of motion for $\theta$ to determine
 them at other times. Alternative, we can solve these two
 equations at each time.

 Going through the above procedure and inverting the equations is
 not straightforward in practice. Let's go over some of the
 cases that we are interested in. Here, we will only invert
 the equations in one space-like slice. This is all we need in order
 to calculate the energies used in section 3. An exactly solvable case is the
 breather. The solution has a special time $t=0$ where the
 sine-Gordon field vanishes. From \nref{sinegordnde} we conclude that $\theta'=0$ at
 $t=0$. Due to the asymptotic boundary conditions we also conclude that
 $\theta = \pi/2$ at $t=0$. We can expand the solution around
 $t=0$ and obtain from \nref{sinegordnde}
 \be \label{breathener}
 \frac{2 \gamma_a}{\cosh \gamma_a x} =
 \frac{\dot{\theta}'}{\sqrt{1-\dot{\theta}^2}}
 \ee
Integrating this equation we find
\be \label{restbre}
 \dot \theta|_{t=0} = { 2 \sinh \gamma_a x \over \cosh^2 \gamma_a
 x }
 \ee
 Using the Virasoro constraints
 \nref{varphidef} we can obtain $\dot{\varphi}$ on the slice,
 which is all we need to calculate the energy.
   The result is the
 expected analytic continuation of the sum of the energy of two
 magnons \nref{bounden} \footnote{Incidentally, this same calculation solves for the energy of
 the, more obvious, scattering solution as well.}.

 Calculating the energy of the boosted breathers is more involved.
 In this case we need to solve the problem around a space-like
 slice $t-v x= 0$. Going through a similar procedure as before
 we get:

 \bea \label{boostener}
 \frac{2 B}{\cosh B u} &=& - \frac{\partial_u^2 \theta}{\sqrt{1-
 (\partial_u \theta)^2}} + \cot \theta \sqrt{1-
 (\partial_u \theta)^2}
 \\ \label{defbandu}
 B &\equiv&\frac{\gamma_a}{v \gamma} ~,~~~~~~~~ u \equiv v\gamma^2(x-vt)
 \eea

  We are interested in solving this equation for boundary conditions such that
   $\theta=\frac{\pi}{2}$ at $u=\pm\infty$, so both endpoints of the string are on the same equator.
    It is important to
  note that the solution to this differential equations depends only on
 $B$ as a parameter. If the sum rule for the energies \nref{bounden} is to hold in
 this case we should obtain
 \be \label{Bformulae}
 E-J=\frac{1}{v} \frac{B}{1+B^2}\\
 \Delta \varphi = 2 p = 4 \arcsin \frac{B}{\sqrt{1+B^2}}
 \ee
Equation \nref{boostener} can be solved explicitly for
$B\rightarrow 0$ and $B\rightarrow \infty$. In both cases we
obtain the expected result. We also solved the problem numerically
(for $B\leq 1$) and found the correct answer. It is most clear,
when solving the problem numerically, that solutions change
character at $B=1$. It is at this point that they reach the pole
of the sphere. Small $B$ solutions are localized in the sphere,
while large $B$ solutions sweep it. Therefore, $B$ is the relevant
parameter to study the type of space-time picture of the solution,
while $q$ controls the period \nref{period}.

Note that if we hold $p$ fixed and we send $q$ or $n$ to infinity,
then we see from \nref{period} that $v \to 0$. The solution to
\nref{boostener} depends on $x$ through a function of $u \sim v x
$ for small $v$. This implies that the size of the state increases
as we increase $n$ keeping $p$ fixed. In fact, it is possible to
check that it increases both in the $x$ and $l$ coordinates, which
are related by \nref{landxen}. For the bound state at rest, large
$n$ means large $a$. We see from \nref{restbre} that its size in
$x$ space is approximately $2a$ while from \nref{bounden} we see
that the energy goes as ${ \sqrt{\lambda} \over \pi} 2 a$. Thus,
from \nref{landxen} we see that the size in the $l$ variables goes
as $\Delta l \sim  2 a { \pi \over \sqrt{\lambda}} $.

\end{document}